\documentclass[twocolumn,showpacs,preprintnumbers,amsmath,amssymb,aps,prl,superscriptaddress]{revtex4-1}
\usepackage{color}
\usepackage{graphicx}
\usepackage{cases}
\usepackage{textcomp} 
\usepackage{subeqnarray}
\usepackage[caption=false]{subfig} 
\begin{document}
\title{\textcolor{black}{Two-phase} flow in a chemically active porous medium}
\author{Alexandre  Darmon}
\affiliation{EC2M, UMR CNRS 7083 Gulliver, PSL Research University, ESPCI ParisTech, 10 Rue Vauquelin, 75005 Paris, France}
\author{Michael Benzaquen}
\affiliation{PCT, UMR CNRS 7083 Gulliver, PSL Research University, ESPCI ParisTech, 10 Rue Vauquelin, 75005 Paris, France}
\author{Thomas Salez}
\affiliation{PCT, UMR CNRS 7083 Gulliver, PSL Research University, ESPCI ParisTech, 10 Rue Vauquelin, 75005 Paris, France}
\author{Olivier Dauchot}
\affiliation{EC2M, UMR CNRS 7083 Gulliver, PSL Research University, ESPCI ParisTech, 10 Rue Vauquelin, 75005 Paris, France}
\date{\today}
\begin{abstract}
We study the problem of the transformation of a given reactant species into an immiscible product species, as they flow through a chemically active porous medium. We derive the equation governing the evolution of the volume fraction of the species -- in a one-dimensional macroscopic description --, identify the relevant dimensionless numbers, and provide simple models for capillary pressure and relative permeabilities, which are quantities of crucial importance when tackling multiphase flows in porous media. We set the domain of validity of our models and discuss the importance of viscous coupling terms in the extended Darcy's law. We investigate numerically the steady regime and demonstrate that the spatial transformation rate of the species along the reactor is non-monotonous, as testified by the existence of an inflection point in the volume fraction profiles. We obtain the scaling of the location of this inflection point with the dimensionless lengths of the problem. Eventually, we provide key elements for optimization of the reactor.
\end{abstract}

\maketitle

Since Darcy's seminal experiments in 1856, flows in porous media have been thoroughly investigated, both at fundamental and industrial levels~\cite{Darcy1856,Muskat1934,Bear1988, Gray2008, Sahimi2011}. A very important area of this research activity concerns the study of two-phase flows for which two fluids in different phases are simultaneously in motion within the same porous medium. The detailed understanding of such two-phase flows is essential to the development of a sustainable economy, as illustrated in hydrology through CO$_2$ sequestration in the soil~\cite{MacMinn2011,Saadatpoor2009}, and groundwater contamination~\cite{Bear2010, Dawson1997, Levy2003}. As of today, the most important industrial application in terms of economical impact remains the extraction technique in the oil industry, which consists in injecting pressurized water in the reservoir to sweep away the hydrocarbon from the rocks~\cite{Satter2008}. Accordingly, most of the studies describing two-phase flows in porous media are carried out in the context of imbibition -- wetting fluid displacing non-wetting fluid~-- or drainage --~non-wetting fluid displacing wetting fluid --, an approach relevant to several but not all situations.

Two-phase flows in complex media raise fundamental questions in hydrodynamics~\cite{Marle1981,deGennes1983,Lenormand1990}, such as the extension of Darcy's law to multiphase flows~\cite{Whitaker1986-1,Whitaker1986-2,Rothman1990}, the importance of viscous coupling~\cite{Kalaydjian1990,Ehrlich1993,Li2005}, as well as the impact of wetting heterogeneities~\cite{Murison2014} and slippage~\cite{Cuenca2013}. Numerous other questions remain open~\cite{Cottin2010,Tallakstad2009}.  
The recent improvement of 3D visualization techniques such as confocal microscopy allows for direct observation of the flow variables and provides access to yet unexplored quantities such as the fluctuations of the flow fields~\cite{Datta2013}, providing answers to old problems and raising new issues. 

If a lot of questions remain unsolved for two-phase flows in inert porous media, this is all the more true when the porous medium becomes chemically active. This is what happens in catalytic reactors, for instance~--~but not only~-- in the petrochemical industry during the cracking and refining processes \cite{Hagen2006}. Those reactors are composed of catalytic grains --~typically zeolite~-- that are compacted and thus create a porous medium in which a mixture of fluids flows and reacts when entering in contact with the catalyzer. The different fluids undergo multiple chain reactions as well as other complex chemical transformations resulting in the production of new fluids, some being immiscible with others. 
In light of the tremendous industrial importance of such processes, it is surprising that the number of theoretical and experimental studies in this context remain scarce.  \textcolor{black}{One of the very few experiments that tackles the above-mentioned problem focuses on the propagation of chemical waves in a porous medium \cite{Atis2013}. However, this study is conducted in the transient regime and thus does not, as we shall see later, correspond to the problem we wish to address. Otherwise,}
most studies combining chemistry and transport in porous media concern mixing and diffusion~\cite{Meheust2013}, where both species are in the same phase. 

The aim of the present \textcolor{black}{article} is to propose mesoscopic and macroscopic descriptions of the flow which takes place in a chemically active porous medium, where fluid phases react and transform into other, possibly immiscible, phases. As we shall see, no good intuition of the phenomenology can be gained from a nai•ve interpolation of the physics at play, neither in \textcolor{black}{non-reactive} two-phase flows inside porous media, nor in monophasic chemical reactors. The purpose of the present study is thus to set a general framework for the new situation we consider, focusing on its general understanding and underlining the new issues raised by its originality. 

\begin{figure}[t!]
\begin{center}
\includegraphics[,scale=0.3]{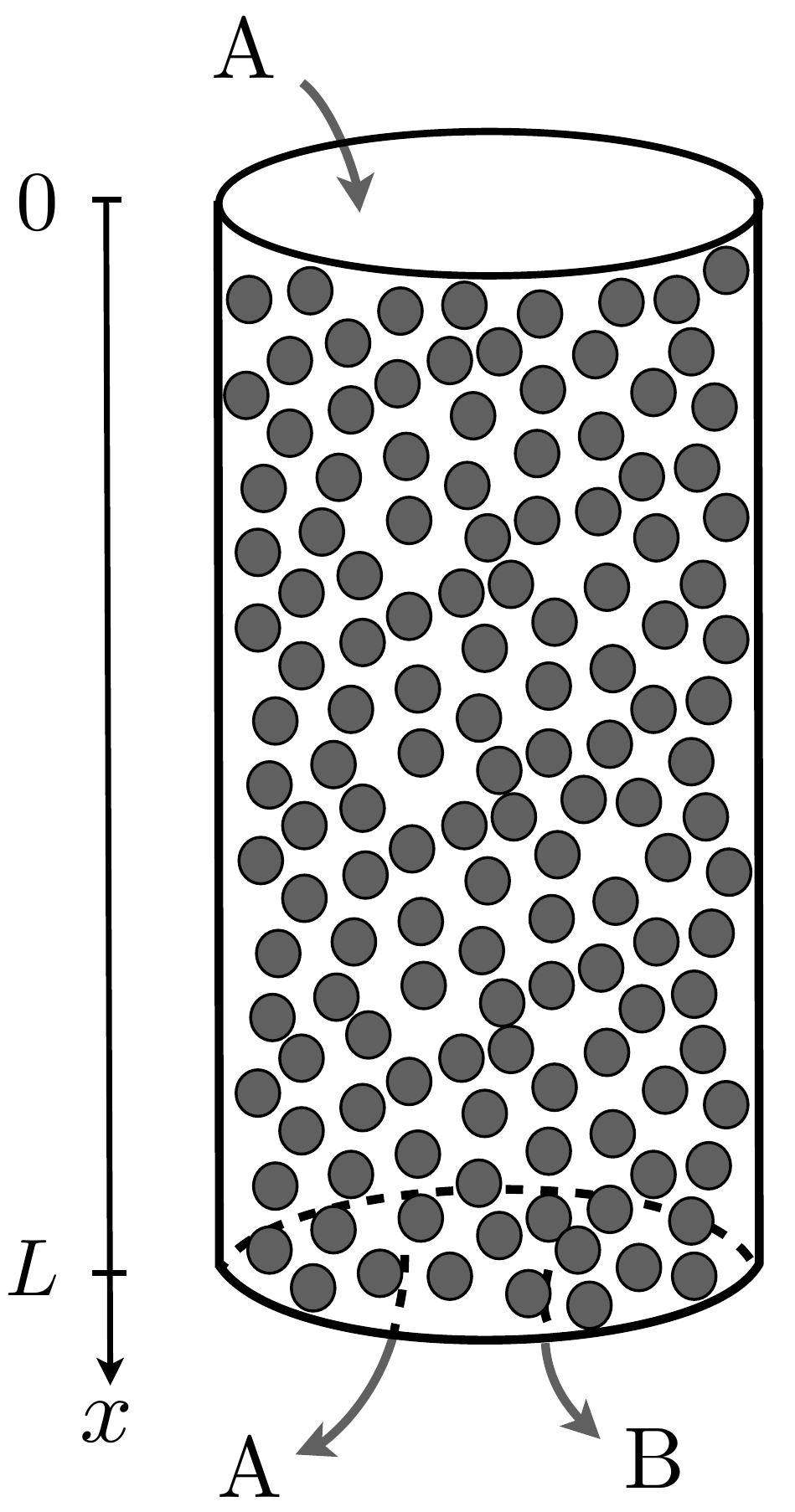}
\caption{Sketch of a catalytic reactor of length $L$ composed of compacted grains. Species A is introduced at the inlet of the reactor ($x=0$) and transformed into species B while flowing inside the porous medium.}
\label{reactor}
\end{center}
\end{figure}

Accordingly, we consider the  simplified situation of a given species A (Fig.~\ref{reactor}), introduced at the inlet of an active porous column, and transformed into another species B, immiscible with species A, as it flows through the reactor. After a phenomenological description of the problem we want to tackle, the comparison is made with simpler natural limit cases of non-reactive two-phase flows and monophasic reactive flows. We also set the frame of description of our problem which corresponds, for the species considered, to a range of volume fractions in which theoretical models can be properly built. We then write down the generalized Darcy's law for this problem and identify the relevant dimensionless parameters. We discuss our choice of relative permeabilities and present a phenomenological model for the capillary pressure. We also highlight the importance of the viscous coupling terms, when extending Darcy's law to two-phase flows, in the absence of which the governing equation is singular. We express a condition for solvability and discuss its relation to Onsager reciprocity relations. Solving the problem numerically along the axis of a unidimensional reactor, we are in position to characterize the evolution of the volume fraction of reactant A along the reactor. We demonstrate two salient effects: the gradient of transformation along the reactor is non-monotonous, as demonstrated by the emergence of an inflection point in the reactant volume fraction profile, and the location of this inflection point exhibits non-trivial scaling with the two dimensionless lengths controlling the problem. Altogether, \textcolor{black}{the proposed model system underlies} 
a complete new class of problems, for which no intuition can be gained from the two natural limit cases: non-reactive two-phase flows and reactive monophasic ones. Finally, we illustrate the principles of a \textcolor{black}{possible} optimization procedure based on the present modelization.

\section{Position of the problem}
In a porous medium, the fluid is moving simultaneously through different \textcolor{black}{interconnected} pores. These pores have different shapes and sizes, making the flow rather complex. Darcy was first to introduce the macroscopic law relating the flow rate $Q$ of a fluid in a porous medium to the pressure gradient $\Delta P/L$ along the medium \textcolor{black}{through}:
\begin{eqnarray}
Q &=& S \frac{\kappa \Delta P}{\mu L}\, , \label{J}
\end{eqnarray}
where $S$ is the cross-sectional area of the medium, $\kappa$ is the permeability, and $\mu$ is the dynamic viscosity of the fluid~\cite{Darcy1856}. The permeability $\kappa$ has the dimension of a length squared and encompasses all the complexity of the porous geometry. It is usually compared to $a^2$, where $a$ is the average pore size, to build up a dimensionless parameter characterizing the porous medium. 

When describing slow flows (no inertial effects) of immiscible fluids, \textcolor{black}{and} neglecting gravity effects, all the physics is controlled by two kinds of mechanisms: viscous pressure drop and \textcolor{black}{capillary pressure}.  
The relative intensities of these two \textcolor{black}{mechanisms} 
are quantified by the capillary number, 
$\textrm{Ca} = \mu V/\gamma$, where $V$ and $\mu$ are respectively the velocity and viscosity of one of the fluids, and \textcolor{black}{where $\gamma$ is} the surface tension between the fluids. The second dimensionless parameter is  the ratio $r$ of the two  
viscosities. In the case of porous media, it was shown that a relevant dimensionless parameter to consider is the extended capillary number $\tilde{\textrm{Ca}} = \textrm{Ca}\times a^2/\kappa$ \textcolor{black}{\cite{Meheust2002,Tallakstad2009}}.  

In this \textcolor{black}{article,} we consider two immiscible fluids, flowing through a porous medium, where a chemical reaction transforms one fluid into the other. We are particularly interested in the evolution of the volume fractions along the medium in the stationary regime. Legitimately, one of the first questions that comes to mind is how the dimensionless numbers are modified by the presence of chemistry. In the simpler situation, where a chemical reaction occurs in a single phase advected with velocity $V$, a natural length scale is $l_\chi = V\tau$, where $\tau$ is the characteristic time scale of the reaction. Identifying the role of this length scale in the case of two-phase flows and expressing how it enters into the relevant control parameters is one of the goals of the present \textcolor{black}{article}.

As mentioned in the introduction, most studies of two-phase flows in porous media deal with  drainage or imbibition~\cite{Bear1988}. In both cases, a given fluid phase is introduced in a porous medium saturated with a second immiscible \textcolor{black}{fluid}  
phase. 
In these situations, a front separating the two species develops and moves along the medium. The propagation modes of this front, its width, and the patterns of which it is composed, severely depend on the capillary number, the viscosity ratio, and in the case of imbibition on the typical pore geometry~\cite{Lenormand1990}.
At the pore level, several scenarii of invasion can take place: selection of the channels depending on their width, trapping of the invading fluid in the smallest channels, etc.
At the macroscopic scale, one looks for a description of the invasion process with a minimal number of variables. Typically, an external pressure gradient or a flow rate is imposed and one monitors the integrated volume fraction, or the longitudinal profile of the fraction of the invading fluid along the porous medium as a function of time.  

In our case, the situation is rather different. 
Let us consider two immiscible fluids A and B flowing in a porous reactor of length $L$ and permeability $\kappa$ (see Fig. \ref{reactor}). While flowing into the porous column, the reactant species A is transformed into the product species B through, for simplicity, a first order chemical reaction characterized by the reaction rate $\tau^{-1}$.  Note that this reaction rate is an effective quantity that depends on the total reactive surface of the pores. In the following, we consider the case where the reactant phase A introduced in the reactor is the wetting phase --~as suggested from the case of cracking encountered in the petroleum industry. The dynamic viscosity of fluid A is denoted $\mu$ and that of B is denoted $r\mu$. Both fluids have the same density, and gravity is not taken into account as the associated pressure gradients are assumed to be negligible compared to those driving the flow.  In the following, $\phi$ denotes the volume fraction of A, and $\phi_{\text B} = 1-\phi$ is that of B. 

\begin{figure}[t!]
\begin{center}
     \includegraphics[,scale=0.2]{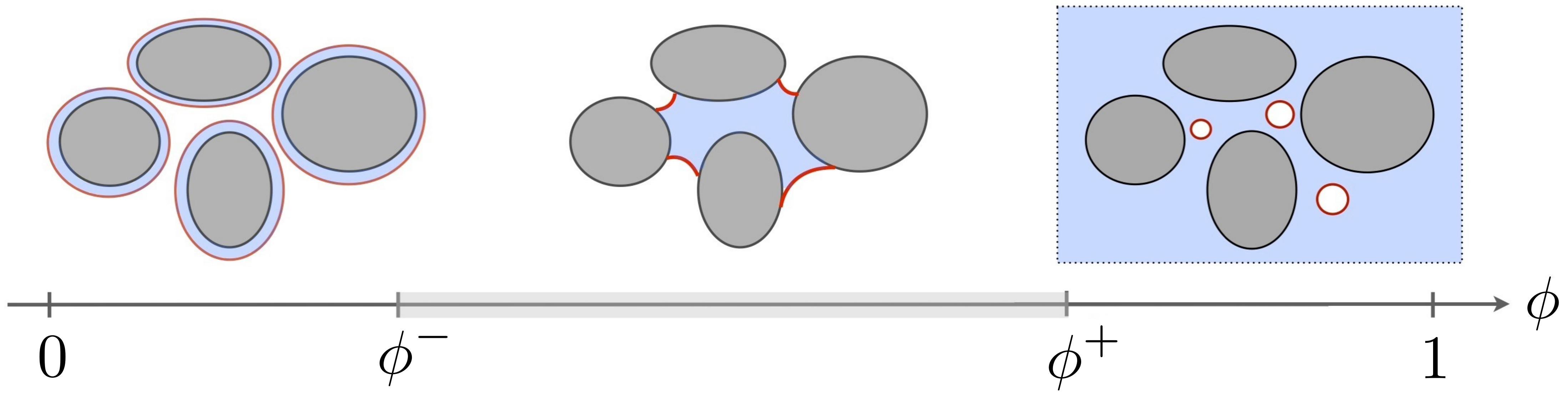}
\caption{Sketch of the three distinguishable regimes when one varies the volume fraction $\phi$ of the wetting fluid. The wetting fluid is signified with blue, the non-wetting fluid is signified with white and the interface between fluids is signified with a red line. For $\phi \in [0,\phi^{-}]$ and $\phi \in [\phi^{+},1]$, \textcolor{black}{only the non-wetting phase and respectively the wetting phase are 
connected in a continuous way.}} 
\label{regimes}
\end{center}
\end{figure}

We are interested in the steady regime, where the reactant phase enters the reactor at an imposed flow rate, and where all the variables are time-independent. Typically, one is interested in quantifying the volume fraction of reactant along the reactor, in order to optimize the flow rate and the reactor length. At first sight, the problem looks simpler than an unsteady non-linear invasion problem.
However, one phase is \emph{permanently} created. Obviously, the inlet of the reactor is a region where droplets of the non-wetting product of reaction B nucleate at the walls and are simply advected by the dominant wetting phase A (see the large $\phi$ regime on Fig.~\ref{regimes}). Similarly, at the outlet of the reactor, when most of the reactant A has been transformed into the non-wetting product of reaction B, the system is full of wetting films along the pore \textcolor{black}{walls} (see the low $\phi$ regime of Fig.~\ref{regimes}).   
\textcolor{black}{These regimes are likely to obey different laws of transport compared to the case where $\phi \in [\phi^{-},\phi^{+}]$. Indeed, in the latter, the transport of matter is mainly driven by displacing  fluid-fluid interfaces within the pores. On the other hand, for the extreme regimes, one phase is continuously connected and the flows, for both species, are mostly driven by advection. It is therefore difficult to model those three regimes in a continuous way.} 
We thus limit our analysis to the reactor core, where $\phi \in \left[\phi^{-}, \phi^{+} \right]$, away from the inlet and outlet. Let us point out the difficulty that will consequently arise when setting the boundary conditions for the integration of the $\phi$ profiles. Let us also anticipate on the analysis to advertise the fact that the physics within the disconnected regions is controlled by the viscous coupling terms, that we will have to consider for the core analysis to be regular.

\section{Modelisation}
\subsection*{Governing equation}
In the porous column, conservation of matter yields: 
\begin{subequations}
\begin{align}
        \frac{\partial \phi}{\partial t} + \boldsymbol\nabla\cdot\boldsymbol{J} _{\text A} + \frac{{\phi}}{\tau} \,\, &= \,\, 0 \label{aa}\\
       \frac{\partial \phi_{\text B}}{\partial t} + \boldsymbol\nabla \cdot\boldsymbol{J}_{\text B} -  \frac{{\phi}}{\tau} \,\, &= \,\, 0 \, , \label{bb}
       \end{align}
       \label{aabb}%
\end{subequations}
\noindent where $\boldsymbol{J} _i$ is the local current of species $i$. Combining Eqs.~\eqref{aa} and \eqref{bb} leads to the conservation of the total current $ \boldsymbol{J}  =  \boldsymbol{J} _{\text A} +  \boldsymbol{J} _{\text B}$:
\begin{eqnarray}
\boldsymbol\nabla\cdot \boldsymbol{J}  &=&0\, . \label{J}
\end{eqnarray}
The extension of Darcy's law to two-phase flows~\cite{Bear2010} reads:
  \begin{subeqnarray}{}
   \boldsymbol{J} _{\text A} (\phi) &=& -\Sigma_{\text A} (\phi) \mathbf{\boldsymbol\nabla}P_{\text A}(\phi)  - \Lambda_{\text{AB}}(\phi) \boldsymbol\nabla P_{\text B}(\phi)  \label{JA}\\
   \boldsymbol{J} _{\text B} (\phi) &=& - \Lambda_{\text{BA}}(\phi) \boldsymbol\nabla P_{\text A}(\phi) -\Sigma_{\text B} (\phi) \boldsymbol\nabla P_{\text B}(\phi)  \, , \label{JB}
\end{subeqnarray}
where $\Sigma_i$ denotes the Darcy permeability of  species $i$, $\boldsymbol\nabla P_i$ is the pressure gradient in the fluid $i$, and $\Lambda_{ij}$ is a viscous coupling term resulting from the action of fluid $j$ on fluid $i$. \textcolor{black}{It is important to note here that Eqs.~\eqref{aabb} and \eqref{JA} correspond to a 3-D mesoscopic description of the problem. Indeed, all the physical quantities present in these equations are already averaged over a Representative Elementary Volume (REV) \cite{Bear1988}.}

In the following, \textcolor{black}{we assume that the pressure and volume fraction gradients in the transverse $y$ and $z$ directions are small compared to those in the $x$-direction along the reactor. This allows us to transform our 3-D mesoscopic equations into a one-dimensional formulation of the problem, where all the physical quantities no longer depend on the $y$ and $z$ positions. For the sake of clarity, we keep the same notations and the bold symbols are replaced by their equivalent scalar expression in the $x$-direction. We denote $x=0$ and
$x=L$ respectively as the entrance and exit of the reactor.} 
In the 1-D \textcolor{black}{description} 
, Eq.~\eqref{J} implies that the total current $J$ is spatially homogeneous. Combining Eq.~\eqref{J} with Eq.~\eqref{JA} leads to:
\begin{eqnarray}
J_{\text A}(\phi) &=& \rho(\phi) J - \Sigma(\phi) \frac{\partial \Pi}{\partial x}\, , \label{JJA}
\end{eqnarray}
where we introduced the pressure difference $\Pi (\phi) = P_{\text A}(\phi) - P_{\text B}(\phi)$, and where we introduced:
\begin{subequations}
\begin{align}
\rho(\phi) &=  \frac{\Sigma_{\text A}(\phi) + \Lambda_{\text{AB}}(\phi) }{\Sigma_{\text A}(\phi) + \Sigma_{\text B}(\phi) + \Lambda_{\text{AB}}(\phi) + \Lambda_{\text{BA}}(\phi)} \,   \label{rho}\\
\Sigma(\phi) &=  \frac{\Sigma_{\text A}(\phi)\Sigma_{\text B}(\phi) - \Lambda_{\text{AB}}(\phi)\Lambda_{\text{BA}}(\phi)}{\Sigma_{\text A}(\phi) + \Sigma_{\text B}(\phi)+\Lambda_{\text{AB}}(\phi)+\Lambda_{\text{BA}}(\phi)}\, . \label{Sigmaphi}
       \end{align}
       \label{SIGMAS}%
\end{subequations}
Combining  Eq.~\eqref{aa} and Eq.~\eqref{JJA} \textcolor{black}{leads} to 
a nonlinear partial differential equation on $\phi$: 
\begin{eqnarray}
\frac{\partial \phi}{\partial t}  - \frac{\partial }{\partial x} \left[\Sigma(\phi) \frac{\partial \Pi}{\partial x} \right] &=& - \, J \frac{\partial }{\partial x} \left[\rho(\phi)\right] - \frac{\phi}{\tau}\, . \label{Cons1}
\end{eqnarray}
Let us now nondimensionalize the problem through:
\begin{eqnarray}
&\displaystyle  \tilde{\Sigma}  = \frac{\mu}{\kappa} \,   \Sigma \,\,\, ; \,\,\,  \tilde{\Pi} = \frac{\Pi}{\gamma / a}  \,\,\, ; \,\,\, X=\frac{x}{L} \, ,\label{ND}
\end{eqnarray} 
where $\gamma$ denotes the surface tension between A and B, and where $a$ is the typical size of the pores. Note that both $\tilde{\Sigma}$ and $\rho$ are functions of the dimensionless variable $r = \mu_{\text B} / \mu$.
In the stationary regime, taking Eq.~\eqref{Cons1} together with Eq.~\eqref{ND} yields the following Painlev{\'e}-like equation:
\begin{equation}
-\xi_1 \left[ F_1(\phi) \, \partial_X^2 \phi +F_2(\phi) \, (\partial_X \phi)^2 \right] + F_3(\phi) \, \partial_X \phi + \frac{1}{\xi_2}\phi = 0,\label{ConsND}
\end{equation}
where:
\begin{subeqnarray}
F_1(\phi) &=& \tilde{\Sigma}(\phi)\, \tilde{\Pi}'(\phi)  \label{A} \\
F_2(\phi) &=&\tilde{\Sigma}(\phi) \,\tilde{\Pi}''(\phi) + \tilde{\Sigma}'(\phi)\, \tilde{\Pi}'(\phi)  \label{B} \\
F_3(\phi) &=& \rho'(\phi) \, , \label{C} 
\end{subeqnarray}
and where the prime denotes the derivation with respect to $\phi$. In addition, we introduced the two dimensionless lengths $\xi_1$ and $\xi_2$:
\begin{subequations}
\begin{align}
\xi_1 &= \frac{1}{\textrm{Ca}} \, \frac{\kappa}{a^2} \frac{a}{L} \,    \label{xi11} \\
\xi_2 &= \frac{l_\chi}{L}  \, ,   \label{xi22}
    \end{align}
       \label{beta1new}%
\end{subequations}
where $\textrm{Ca} = \mu J / \gamma$ is the usual capillary number, and where $l_\chi = J\tau$ is the typical length over which the chemical reaction occurs. Hence, varying $\xi_1$ and $\xi_2$ allows to control independently the effects of visco-capillarity and chemistry. Altogether, the three relevant dimensionless parameters to describe the problem are thus $\xi_1$, $\xi_2$ and $r$. Note that comparing the local visco-capillarity and chemistry, one can get rid of $L$ in Eqs.~\eqref{beta1new} and obtains:
\begin{equation}
\frac{\xi_2}{\xi_1} = \textrm{Ca} \frac{a^2}{\kappa} \frac{l_\chi}{a} \, .    \label{beta} 
\end{equation}
This shows that, when adding chemistry to two-phase flows in porous media, the usual dimensionless number $\tilde{\textrm{Ca}} = \textrm{Ca} \times a^2/\kappa$ \textcolor{black}{has to} 
be renormalized by a factor $l_\chi /a$\textcolor{black}{, also often referred to in the literature as the Damk\"ohler number \cite{Meheust2013}.} \\

To proceed further, one must  prescribe functional forms for the relative permeabilities $\Sigma_{\text A}$ and $\Sigma_{\text B}$, the viscous coupling coefficients $\Lambda_{\text{AB}}$ and $\Lambda_{\text{BA}}$, and the pressure difference $\Pi$.
\textcolor{black}{Even though studies have been conducted and have succeeded in describing with accuracy the mechanisms at stake at the pore scale, the description at the level of the REV is still not completely understood and does not allow for a direct derivation of the above-mentioned mesoscopic quantities.} 
One should thus, in principle, rely on experimental observations \textcolor{black}{and multi-scale numerics \cite{Geiger2004}}. However, even in the case of drainage and imbibition, for which \textcolor{black}{numerous} experimental studies exist, there is not yet a general consensus on what should be the exact form of these functions, one reason being that the quantities measured experimentally are usually not directly related to those functions. In the case of chemical reactors, there is even less experimental data. In the following, we will thus concentrate on the general trends of these functions in terms of variation and concavity. We will use simple analytic forms for the purpose of the analysis, but these should not be considered as realistic models. 

\subsection*{Relative permeabilities}
Let us first look into the relative permeabilities $\Sigma_{\text A}$ and $\Sigma_{\text B}$, which are the self-response coefficients of the flow of species $i$ to a pressure gradient within the same phase $i$. They have been extensively studied as they are of crucial importance in the description of these complex \textcolor{black}{multiphase} flows. Different laws of evolution have been proposed, both phenomenological and theoretical \cite{Burdine1952,Brooks1964,Bear1988}. But there is no unified theory and there only seems to be a general agreement on their qualitative shape. In the regimes where $\phi \in [0,\phi^{-}]$ and $\phi \in [\phi^{+},1]$, the diagonal coefficients $\Sigma_\text A$ and $\Sigma_\text B$ are equal to 0 respectively since the flow of the minority species is almost entirely due to the pressure gradient in the majority phase. In the range $[\phi^{-}, \phi^{+}]$, $\Sigma_\text A$ increases as the volume fraction $\phi$ of A increases. This can be easily understood by noting that for a given species, the overall effective permeability of the porous medium somehow comprises the volume occupied by the other species because of immiscibility. In fact, as $\phi$ becomes larger, B occupies less volume and thus more free volume is accessible to A. Similarly, the relative permeability of B decreases as $\phi$ increases. Again, as we are not interested here in the specific form of the relative permeabilities, but rather on their evolution trends with volume fraction, we take a simple phenomenological power-law form for $\phi \in [\phi^-, \phi^+]$, inspired by the shape they take in the literature \textcolor{black}{\cite{Brooks1964}}:
\begin{subeqnarray}
\Sigma_\text A (\phi) &=& \alpha_\text A \frac{\kappa}{\mu} \left(\frac{\phi - \phi^-}{\phi^+ - \phi^-}\right)^{\nu} \, \label{sigmaAphi}\\
\Sigma_\text B (\phi) &=& \alpha_\text B \frac{\kappa}{r \mu} \left(1-\frac{\phi - \phi^-}{\phi^+ - \phi^-}\right)^{\nu} \, . \label{sigmaBphi}
\end{subeqnarray}
where $\nu$, $\alpha_\text A$ and $\alpha_\text B$ are positive \textcolor{black}{dimensionless} numbers.

\subsection*{Viscous coupling}
The evolution of the viscous coupling coefficients $\Lambda_{ij}$  with the volume fraction has been a matter of debate \cite{Kalaydjian1990,Ehrlich1993,Avraam1995}, and no clear tendency has yet been highlighted. \textcolor{black}{An interesting example of experimental evaluation of the evolution of the viscous coupling terms consists in studying the flow of a wetting fluid trapped at the corners of a pore throat \cite{Du2012}. The flow in the wetting phase is due to a pressure gradient in the non-wetting phase and one has then access to the viscous coupling coefficient at the pore scale. Nevertheless, in most problems tackling two-phase flow in porous media, the viscous coupling terms }are  set to zero \cite{deGennes1983}, to simplify calculations and under the justification that, when the contact angle of the interface on the solid matrix is finite, there is mostly no interface parallel to the streamlines at the pore scale. However, as we shall see later, they are of crucial importance and their presence is necessary to solve the present problem. Viscous couplings can be seen as flow response coefficients to a pressure gradient in the other phase. When $\phi<\phi^{-}$ or $\phi>\phi^{+}$, the flow of the discontinuous phase is almost entirely due to pressure gradients in the continuous one. 
Therefore, the viscous coupling coefficients $\Lambda_{\text{AB}}$ and $\Lambda_{\text{BA}}$ are respectively non-zero in these ranges of volume fraction. \textcolor{black}{For $\phi \in [\phi^{-},\phi^{+}]$, both phases play symmetrical roles in terms of hydrodynamics. Therefore, the viscous coupling coefficients should have similar dependencies on $\phi$ and should only differ in magnitude because of the different viscosities of the two phases; in the range $\phi \in [\phi^{-},\phi^{+}]$, one expects $\Lambda_{\text{AB}}(\phi) = r \Lambda_{\text{BA}}(\phi)$. Let us now specify two interesting values for the viscous coupling coefficients, namely $\Lambda_{\text{AB}}(\phi^-)$ and $\Lambda_{\text{BA}}(\phi^+)$.}  
When $\phi>\phi^{+}$, the disconnected phase is composed of droplets advected by the continuous \textcolor{black}{phase}, while for $\phi<\phi^{-}$, the disconnected phase is made of films wetting the porous medium (see Fig. \ref{regimes}). Since isolated droplets are more easily advected than films wetting on the porous medium, 
one expects \textcolor{black}{$\Lambda_{\text{BA}}(\phi^+)>\Lambda_{\text{AB}}(\phi^-)$, which we then parametrize by $\Lambda_{\text{AB}}(\phi^-)=(1-\zeta)\Lambda_0$ and $r \Lambda_{\text{BA}}(\phi^+)=(1+\zeta)\Lambda_0$, where $\Lambda_0$ is the typical viscous coupling, and $\zeta$ is the relative distance to the average taken by the extreme values at $\phi=\phi^-$ and $\phi=\phi^+$}. 

\textcolor{black}{Consistently with the previous discussion, }there is thus no reason to set the viscous coupling to zero, which would require unphysical and non-motivated arguments. \textcolor{black}{The simplest formulation here is to draw a linear interpolation for $\Lambda_{\text{AB}}(\phi)$ and $\Lambda_{\text{BA}}(\phi)$ that reads:
\begin{eqnarray}
\Lambda_\text{AB}(\phi) =r \Lambda_\text{BA}(\phi)= m \phi +p \, , \label{Lambda}
\end{eqnarray}
with: 
\begin{subequations}
\begin{align}
m &=  \frac{2 \ \zeta \, \Lambda_0}{\phi^+-\phi^-} \\
p &= \Lambda_0\left(1- \zeta\,  \frac{\phi^+ + \phi^-}{\phi^+-\phi^-}\right) \, .
    \end{align}
       \label{mp}%
\end{subequations}}

\subsection*{Capillary pressure}
The last key element to prescribe in order to achieve a complete description is the average capillary pressure, \textit{i.e.} the positive pressure difference between non-wetting and wetting phases, $P_\text{B} - P_\text{A} = -\Pi$. The average pressure in each phase is defined as $P_i = \left<p_i\right>$, where $p_\text{i} $ is the pressure in fluid $i$ defined at the pore scale, and where the average is performed over the \textcolor{black}{REV}. 
At the pore scale, the origin of the capillary pressure is the pressure discontinuity at the interface between the two fluids due to the total curvature $\mathcal{C}$ of the interface, following Young-Laplace's law: $\Delta p=\gamma \mathcal{C}$. \\

Several studies have been conducted in order to get data for the capillary pressure as a function of the volume fraction of the wetting fluid~\cite{Leverett1941,Rose1949,Purcell1949,Brown1951}. However, in most experiments the  pressure which is measured is the external pressure drop needed to displace the fluids. This, in principle, can be very different from the above-mentioned capillary pressure $P_\text{B} - P_\text{A}$, which is defined locally~\cite{Bear1988}, and thus the use of the term \textit{capillary pressure} to describe the reported experimental curves can therefore be questioned~\cite{Hassanizadeh1993}.  

\begin{figure}[t!]
\begin{center}
     \includegraphics[,scale=0.35]{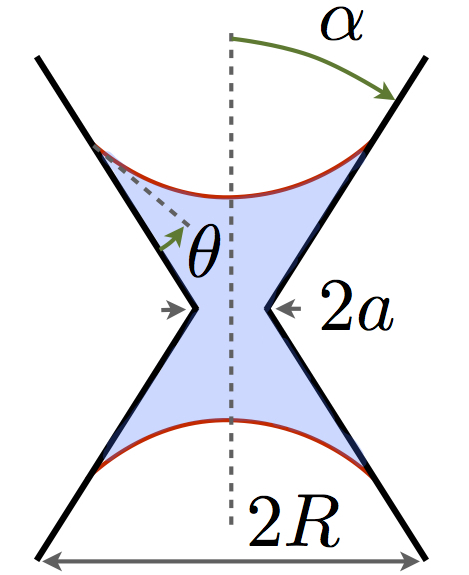}
\caption{Sketch of a single idealized pore with a conical shape.}
\label{Paramcone}
\end{center}
\end{figure}

Modeling the complex geometry of the pores is always a difficult task \textcolor{black}{\cite{Geiger2004}} and is far beyond the scope of this \textcolor{black}{article}. Simple approaches have been proposed: interconnected cylinders, compacted beads, etc.~\cite{Collins1961}.  \textcolor{black}{Somehow similar to the hourglass model of \textit{Aker et al.} \cite{Aker1988},} we here propose a model of connected cones (see Fig. \ref{Paramcone}), as suggested by the intuition of a packing of hard spheres. Yet simple, this toy model is sufficient to take into account the essential idea that the average curvature of the interface between A and B changes with the volume fraction, therefore affecting the local capillary pressure. This crucial element would not appear in a model of connected cylinders of equal sizes. 
Within this simple geometrical model (see Fig. \ref{Paramcone}), \textcolor{black}{one can write down the local capillary pressure as $p_\textrm{c} = p_\text B - p_\text A = 2\gamma/r_\text c(\phi)$, where $r_\text c$ is the curvature radius which directly depends on the volume fraction $\phi$ of the wetting fluid A. Relating $r_\text c$ to the other geometrical parameters then yields:} 
\begin{eqnarray}
p_\textrm{c}(\phi) = \frac{\gamma \, \cos (\alpha +\theta)}{(R^3-a^3)^{1/3}} \, \,C_1 \,(\phi+C_2)^{-1/3} \, ,
\label{Pi}
\end{eqnarray}
where $C_1$ and $C_2$ are positive coefficients that depend on $\alpha, \theta, a$ and $R$ (see Fig.~\ref{Paramcone}). \textcolor{black}{Remember that our aim is to characterize the capillary pressure $P_\text{B} - P_\text{A}$ in the REV, at the mesoscopic scale, where $P_\text{A}$ and $P_\text{B}$ appear in the extended Darcy's law of Eq.~\eqref{JA}. To do so, one has to perform an upscaling step which is still today, as mentioned previously, one of the most intricate point in the theory of porous media \cite{Bear1988}.} 
Within the porous medium, the pores are of course non-identical and the parameters $a, R$ and $\alpha$ vary from one pore to the other. \textcolor{black}{Rigorously, one should therefore compute the upscaling operation to obtain the mesoscopic capillary pressure. Here, given the absence of singularity in the dependence of the local capillary pressure on the volume fraction of wetting fluid, we assume}
 a simple renormalization of these constants when averaging over the distribution of pores --~a reasonable assumption if one considers homogeneous wettability conditions~--, \textcolor{black}{and} make the hypothesis that the average capillary pressure \textcolor{black}{in the REV} follows: 
 \begin{eqnarray}
P_\text{B} - P_\text{A} &=& -\Pi = \frac{\gamma}{a} D_1 \, (\phi + D_2)^{-1/3} \, , \label{Mic}
\end{eqnarray}
where $D_1$ and $D_2$ are positive constants depending on the distributions of angles and pore sizes. Note that similar power-law dependencies of the capillary pressure with volume fraction have been derived previously~\cite{Brooks1964}, the main difference being that the exponent therein depends on the shape of the pore distribution, whereas here the $-1/3$ exponent is a direct signature of 3D volumic effects. 

\textcolor{black}{\subsection*{Summary}}

\begin{figure}[t!]
\begin{center}
     \includegraphics[,scale=0.183]{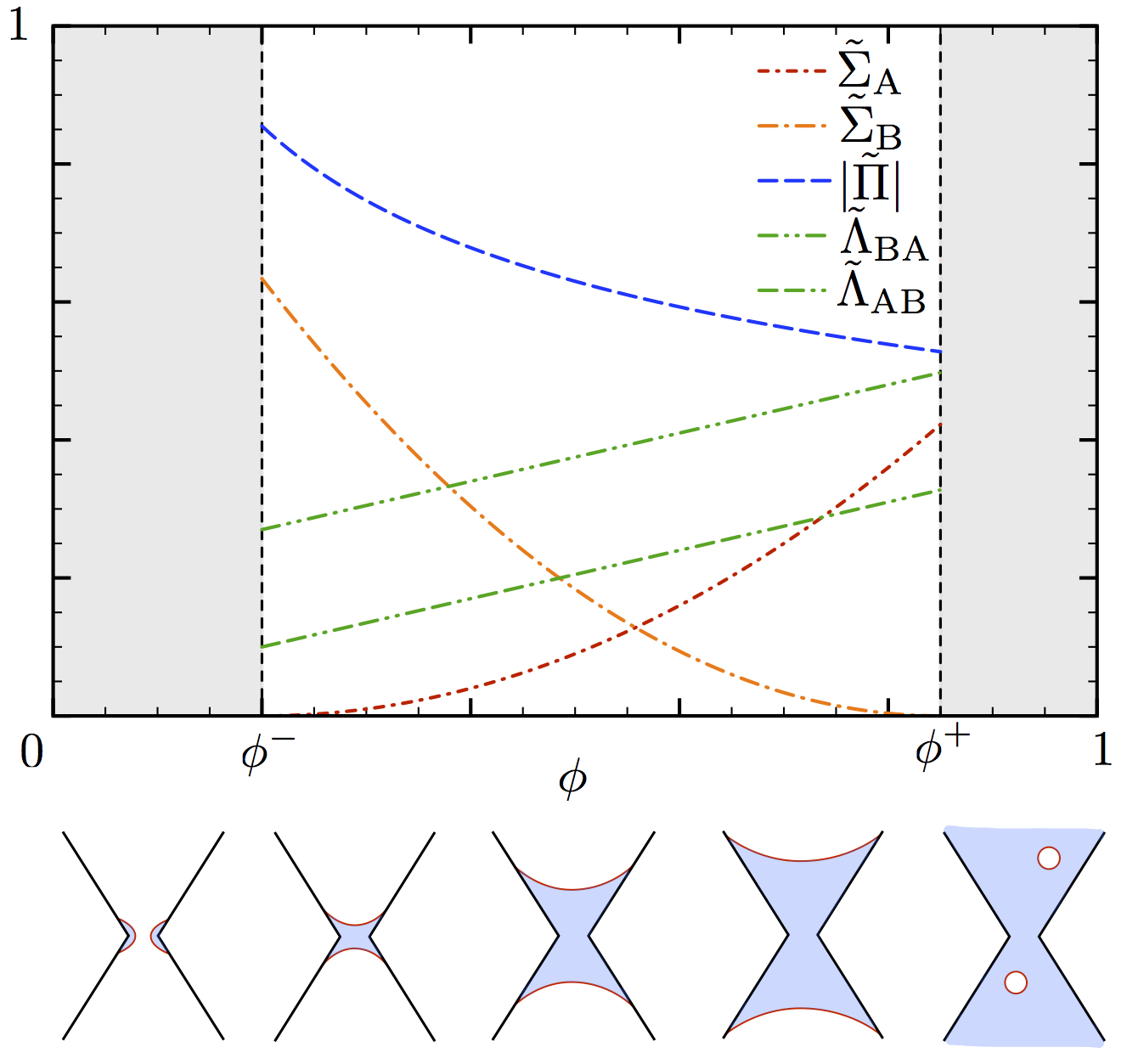}
\caption{Typical behavior for the dimensionless average capillary pressure $|\tilde{\Pi}|$ (blue),  
dimensionless relative permeabilities ${\tilde{\Sigma}_{\text A}}$ (red) and ${\tilde{\Sigma}_{\text B}}$ (orange), and dimensionless viscous couplings \textcolor{black}{$\tilde{\Lambda}_\text{AB}$ and $\tilde{\Lambda}_\text{BA}$} 
(green), as a function of the volume fraction $\phi$ of the wetting fluid A.}
\label{resume}
\end{center}
\end{figure}

Altogether, we end up with simple functional laws for the permeabilities, the viscous coupling \textcolor{black}{coefficients} and the capillary pressure, as a function of the volume fraction of the wetting phase.
Figure~\ref{resume} offers a synthesis of these dependencies in the range $[\phi^-, \phi^+]$.  
The functional forms given by Eqs.~\eqref{sigmaAphi}, \eqref{Lambda} and \eqref{Mic}, together with Eq.~(\ref{ConsND}), define the equation to be solved. 

\medskip

\section*{Resolution}

We are now in position to solve Eq.~(\ref{ConsND}), provided that we have two boundary conditions at the inlet of the integration domain. One must remember that the description we proposed is valid only when $\phi \in [\phi^-, \phi^+]$. Hence, at the entrance of the resolution domain, $X=0$, we can set the volume fraction of wetting fluid A, $\phi(X=0) =\phi^+$, but not larger. A natural second boundary condition is to impose the continuity of $\partial_X \phi$ in $X=0$. We choose to describe the situation where the negative $X$ domain describes the same porous media, albeit inactive --~so that chemical reactions do not take place~-- but in which care has been taken to inject the two phases A and B with $\phi(X\le0) = \phi^+$. We thus have $\partial_X \phi (X=0) = 0$. In practice, the above strategy consists in having a first layer of passive porous media at the entrance of the chemical reactor, and injecting both fluids with a small but finite amount of the product of the reaction.

To simplify the notations, it is now convenient to introduce a new field: 
\begin{eqnarray}
\psi(X) = \frac{\phi(X)-\phi^-}{\phi^+-\phi^-}\, , \label{psi}
\end{eqnarray}
and rewrite the governing Eq.~\eqref{ConsND} in terms of $\psi$ as: 
\begin{equation}
-\xi_1[A(\psi) \, \partial_X^2 \psi+B(\psi)\,(\partial_X \psi)^2] + C(\psi) \,\partial_X \psi+ \frac{\psi}{\xi_2}= \frac{K}{\xi_2}, \,\,\,\,\,\,\,\,\,\, \label{ConsNDpsi}
\end{equation}
with:
\begin{subequations}
\begin{align}
A(\psi) &=  \frac{1}{\phi^+-\phi^-} \widehat{\Sigma}(\psi)\, \widehat{\Pi}'(\psi) \,\,\,\,\,\,  \label{A} \\
B(\psi) &=\frac{1}{\phi^+-\phi^-} \left[\widehat{\Sigma}(\psi) \,\widehat{\Pi}''(\psi) + \widehat{\Sigma}'(\psi)\, \widehat{\Pi}'(\psi)\right] \,\,\,\,\,\,  \label{B} \\
C(\psi) &= \frac{1}{\phi^+-\phi^-} \, \widehat{\rho} \, '(\psi) \,\,\,\,\,\,  \label{C} \\
K &= -\frac{\phi^-}{\phi^+-\phi^-}  \label{D}  \,\,\,\,\,\, ,
\end{align}
\label{AABB}%
\end{subequations}
and the boundary conditions:
\begin{subequations}
\begin{align}
\psi(0)&=1 \, \label{psi0} \\
\partial_X \psi(0)&=0 \, , \label{psi'0}
\end{align}
\label{boundary}%
\end{subequations}
where the prime now denotes the derivation with respect to $\psi$, and where: $\widehat{\Sigma}(\psi) = \tilde{\Sigma}(\phi)$, $\widehat{\Pi}(\psi) = \tilde{\Pi}(\phi)$, and $\widehat{\rho}(\psi) = \rho(\phi)$, for $\phi \in [\phi^{-}, \phi^{+}]$.
Equations~\eqref{ConsNDpsi}, \eqref{AABB} and \eqref{boundary} describe our problem. In the following, we discuss the solvability condition for Eq.~\eqref{ConsNDpsi} and solve it numerically, after an analytical study of two important limit cases. 

\subsection*{Solvability condition}

A necessary condition required to solve numerically Eq.~\eqref{ConsNDpsi}  is that the coefficient in front of the term of highest differential degree, that is $A(\psi)$ given by Eq.~\eqref{A}, does not cancel out. From Eqs.~\eqref{Mic} and \eqref{psi}, we get:
 \begin{eqnarray}
\widehat{\Pi}(\psi) &=& - \widehat{\Pi}_0 \, (\psi + \widehat{D})^{-1/3} \, , \label{Pipsi}
\end{eqnarray}
where $\widehat{\Pi}_0 = D_1 (\phi^{+}-\phi^{-})^{1/3}$,  and $\widehat{D} =  (D_2+\phi^{-})/(\phi^{+}-\phi^{-})$. We see that $\widehat{{\Pi}}'(\psi)>0$, for all $\psi \in [0,1]$. The sign of $A(\psi)$ is then given by that of $\widehat{\Sigma}(\psi)$. Recalling Eq.~\eqref{Sigmaphi} and given that all Darcy coefficients are positive, the solvability condition is therefore restricted to the study of the sign of $\widehat{\Sigma}_{\text A}(\psi)\widehat{\Sigma}_{\text B}(\psi) - \widehat{\Lambda}_{\text{AB}}(\psi)\widehat{\Lambda}_{\text{BA}}(\psi)$, with $\widehat{\Sigma}_i (\psi)= \tilde\Sigma_i(\phi)$ and $\widehat{\Lambda}_{ij} (\psi)= \tilde\Lambda_{ij}(\phi)$. 
The boundary condition $\psi(X=0) =1$ imposes:
\begin{eqnarray}
\widehat{\Sigma}_{\text A}(1)\widehat{\Sigma}_{\text B}(1) - \widehat{\Lambda}_{\text{AB}}(1)\widehat{\Lambda}_{\text{BA}}(1) = - \textcolor{black}{r \Lambda_\text{BA}(\phi^+)^{2}}
 < 0 \, .  \,\,\,\,\,\,\,\,\,\,
\end{eqnarray}
Since for numerical solvability $\widehat{\Sigma}_{\text A}(\psi)\widehat{\Sigma}_{\text B}(\psi) - \widehat{\Lambda}_{\text{AB}}(\psi)\widehat{\Lambda}_{\text{BA}}(\psi)$ must not change sign along the reactor, we obtain the following solvability condition on Darcy's law coefficients:
 \begin{eqnarray}
\widehat{\Lambda}_{\text{AB}}(\psi)\widehat{\Lambda}_{\text{BA}}(\psi)\, > \, \widehat\Sigma_{\text A}(\psi)\widehat\Sigma_{\text B}(\psi)  \ ,\label{antiONS}
\end{eqnarray}
for all $\psi \in [0,1]$. \textcolor{black}{To gain more physical insight on this solvability condition, we choose, as it will be the case for the numerical resolution, the simplest functional form for the viscous coupling terms and set $\zeta=0$ such that $\Lambda_\text{AB}(\phi^-)=\Lambda_0=r\Lambda_\text{BA}(\phi^+)$}.  
With the specific functional form chosen for the permeabilities as given by Eq.~\eqref{sigmaAphi}, \textcolor{black}{and by introducing the dimensionless viscous coupling coefficient $\tilde{\Lambda}_0=\Lambda_0 \mu/\kappa$, }the solvability condition \textcolor{black}{then} summarizes into:
\begin{eqnarray}
\textcolor{black}{\tilde{\Lambda}_0 \, > \, \sqrt{\alpha_\text A \alpha_\text B} ~\left(\frac{1}{2}\right)^{\nu} } .\label{antiONS2} 
\end{eqnarray}
Not only \textcolor{black}{must} the viscous coupling terms be non-zero for the problem to be solvable, but they must be larger than a finite constant imposed by the permeabilities.  
This is a strong condition, without which singularities set in.  It is interesting to note that setting the viscous coefficients to zero, as often done in the literature for simplifications \cite{deGennes1983}, does not fulfill the solvability condition. Here, the singularities are revealed by the terms in $1/\xi_2$ of Eq.~\eqref{ConsNDpsi}, associated with the chemistry, and as such they would not appear for a standard two-phase flow problem. However, the solvability condition found here is a condition on the non-vanishing coefficient for the largest derivative order term, which therefore fixes the nature of the mathematical problem to be solved. This observation suggests that reconsidering the standard two-phase flow problems with non-vanishing viscous coupling terms would possibly lead to yet unexplored and qualitatively new flow regimes.

It is also worth mentioning that, at first sight, Eq.~\eqref{antiONS} appears as the exact opposite inequality compared to Onsager's reciprocity relation~\cite{degroot2013}. The applicability of Onsager's theory of near-equilibrium systems to two-phase flows in porous media has itself been another important matter of debate \cite{Kalaydjian1990,Gunstensen1993,Bentsen1994,Avraam1995}. Let us only mention here that writing properly the fluxes and affinities for the two-phase flow problem, we find that Onsager's inequality does not directly apply for the very coefficients of Eqs.~\eqref{JA}. The reason is that, apart from the additional role of chemistry, the true affinities are not the pressure gradients but rather the generalized chemical potential gradients which include the effects of hydrodynamics and, in our case, chemistry. The applicability of Onsager's reciprocity relation would therefore actually hold for a new formulation of the extended Darcy's law with the correct thermodynamic fluxes and corresponding affinities. 

\section*{Solutions and Optimization}

Before solving the full two-phase flow problem with chemistry, let us briefly discuss the simpler limiting cases of a two-phase flow without chemistry, and that of a monophasic flow with chemistry \textcolor{black}{where chemistry does not change viscosity ($r=1$).} 

\subsection*{Two-phase flow without chemistry}

As mentioned previously, most studies conducted for two-phase flows in porous media, without any chemistry involved, focus on time-dependent evolutions and mostly in the case of imbibition or drainage. Only very few investigations have been led in the past on the steady-state of two immiscible fluids flowing simultaneously. Tallakstad \textit{et al.}~\cite{Tallakstad2009} conducted a quasi-2D experimental study of the steady-state, in the case where isolated clusters of one phase co-exist with a continuous phase, and characterized the size distribution of these clusters. Namely, they investigated the \textcolor{black}{extreme regimes} which the present continuous formulation can not describe. As a matter of fact, the present formulation turns out to have a trivial steady behavior. Eqs.~\eqref{aabb} simply become $ \boldsymbol\nabla\cdot\boldsymbol{J} _{\text A} =  \boldsymbol\nabla\cdot \boldsymbol{J} _{\text B}  = 0$. The flow of each species is therefore spatially homogeneous; there is no evolution of the volume fractions along the reactor and the solution is $\psi (X) = \text{Const.} = \psi(0)$. 

\subsection*{Monophasic flow with chemistry}

Consider now the case where A and B are miscible\textcolor{black}{, with same viscosities ($r=1$),} and only one common phase flows into the reactor. \textcolor{black}{In the stationary regime,} conservation of matter yields:
\textcolor{black}{\begin{equation}
\frac{\text d J_\text A}{\text d x} + \frac{\phi(x)}{\tau} = 0 \, . \label{expphix}
\end{equation}
Noting that $J_\text A(x)=J \phi(x)$, Eq. \eqref{expphix} can then be rewritten in terms of $\psi$ and $X$ as:}
\begin{equation}
\frac{\text d \psi}{\text d X} + \frac{\psi(X)}{\xi_2} = \frac{K}{\xi_2} \, , \label{exp}
\end{equation}
where $\xi_2$ and $K$ are defined in Eqs.~\eqref{xi22} and \eqref{D}. The solution to Eq.~\eqref{exp} is an exponential function that reads:
\begin{equation}
\psi(X) = [\psi(0)-K]\,e^{-X/\xi_2} + K \, .
\end{equation}
where $X/\xi_2 = x / J\tau$. The two species are simply advected with a constant current $J$, on the length scale $l_\chi = J\tau$, while the chemical reactions turns A into B with the timescale $\tau$. 

\subsection*{General case: numerical solutions}

Considering now the general two-phase case with chemistry, it is hard to gain any intuition from the two above limiting cases, which happen to be far too simple. Mathematically, this corresponds to the fact that, by coupling chemistry and capillarity, the governing equation becomes highly nonlinear and of second order. \textcolor{black}{At this stage, it is worth mentioning that the dimensionless parameters can be classified into two different categories. The first one corresponds to the parameters that characterize the problem of a two-phase flow in a chemically active porous medium, namely $\xi_1$, $\xi_2$ and $r$. The aim is thus to perform a parametric study, and identify different behaviors for the evolution of the volume fraction when varying those parameters. The other parameters such as $\alpha_\text A, \alpha_\text B, \nu, \tilde{\Lambda}_0, \zeta$ are model dependent. A complete parametric study of the model is far beyond the scope of the present work. We therefore consider the simplest functional forms for the models previously introduced. We let $\nu=1$ in Eq.~\eqref{sigmaAphi}. Indeed, taking linear evolutions of the relative permeabilities is sufficient to describe their qualitative variation with the volume fraction.  Consistently with this first order approach, we then set $\zeta=0$, $\tilde{\Lambda}_0 = 1$, and $\alpha_\text A = \alpha_\text B = 1$. We checked that other reasonable values of $\nu$ or any other model dependent parameters lead to similar conclusions. Finally, since we considered a simple chemical transformation A $\rightarrow$ B, with no stoichiometry involved, the viscosity ratio satisfies $r \sim 1$, and thus $r=1$ will be taken as a typical reference value throughout the following. Note finally that the values chosen for the coefficients fulfill the solvability condition given by Eq.~\eqref{antiONS2}. }

\begin{figure}[t!]
\begin{center}
     \includegraphics[,scale=0.47]{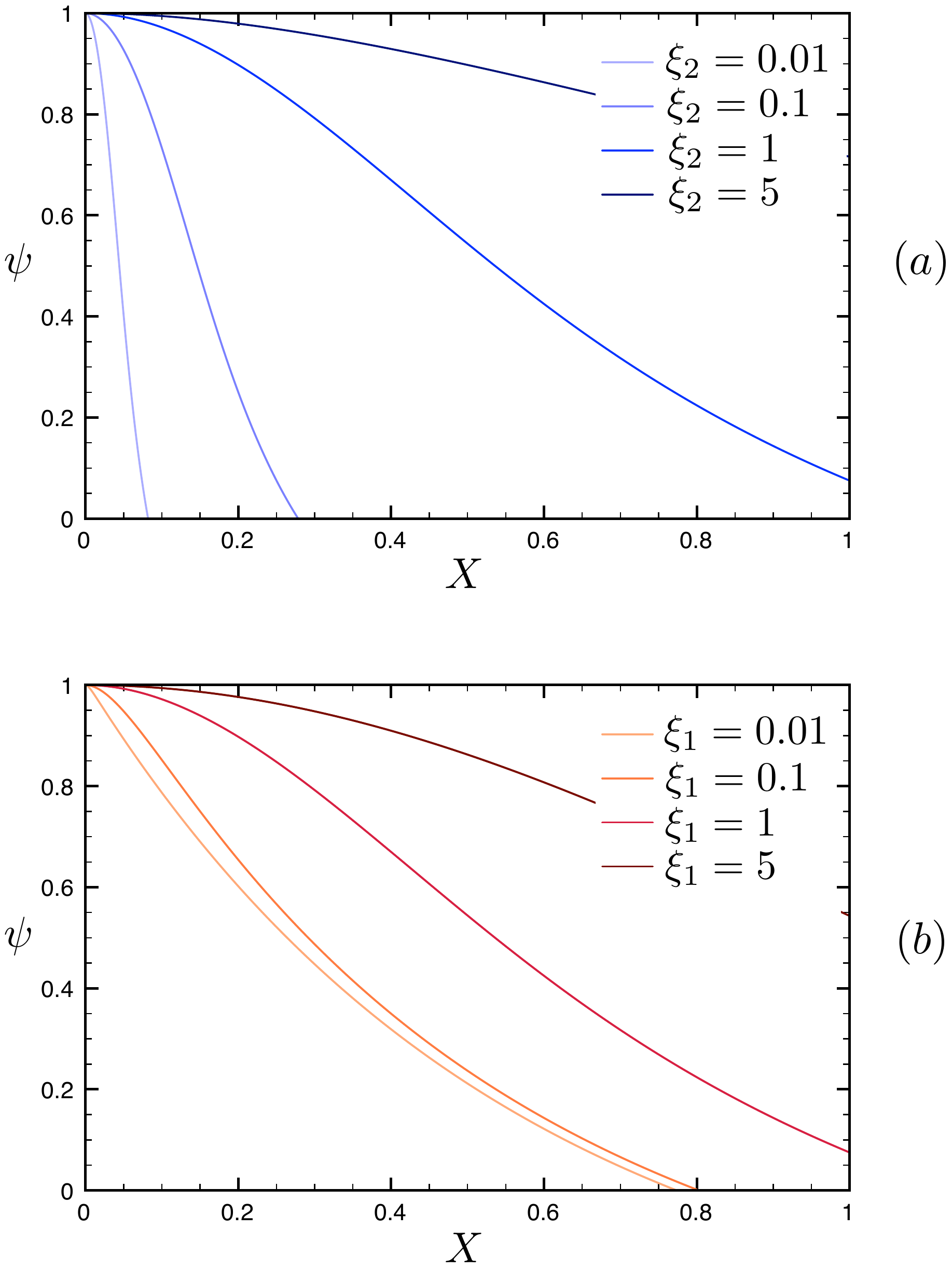}
\caption{Renormalized volume fraction $\psi$ against the dimensionless position $X$ in the reactor,  for different sets of parameters ($\xi_1$, $\xi_2$), obtained from numerically solving Eq.~\eqref{ConsNDpsi}. The parameters $\tilde{\Lambda}_0$, $\alpha_\text A$, $\alpha_\text B $ and $r$ are set to 1. (a) $\xi_2$ varies and $\xi_1$ is fixed to 1. (b) $\xi_1$ varies and $\xi_2$ is fixed to 1.}
\label{solutions1}
\end{center}
\end{figure} 

We use a Runge-Kutta algorithm \cite{Press1988} to solve numerically Eq.~\eqref{ConsNDpsi}, with the phenomenological models built in Eqs.~\eqref{sigmaAphi}, \eqref{Lambda} and \eqref{Mic}, and the boundary conditions of  Eq.~\eqref{boundary}. 
Figure~\ref{solutions1} displays plots of the renormalized volume fraction $\psi$ against the dimensionless position $X$ in the reactor, for two different sets of control parameters ($\xi_1$, $\xi_2$). In Fig.~\ref{solutions1}(a), $\xi_1$ is fixed to 1 and $\xi_2$ is varied as indicated, whereas in Fig.~\ref{solutions1}(b), $\xi_2$ is fixed to 1 and $\xi_1$ is varied.
The first significant observation is that, in all cases, the volume fraction profiles are significantly distinct from a simple exponential relaxation along the reactor. In particular, at the beginning of the reactor, the convexity is reversed, leading to the existence of an inflection point, at \textcolor{black}{a} finite distance from the origin. The volume fractions do not evolve significantly at the entrance of the reactor; this is then followed by a more pronounced decrease, where most of the chemical transformation occurs on a relatively small length. Finally, the curvature changes sign and the evolution takes place on larger length scales. 

\begin{figure}[t!]
\begin{center}
\includegraphics[,scale=0.6]{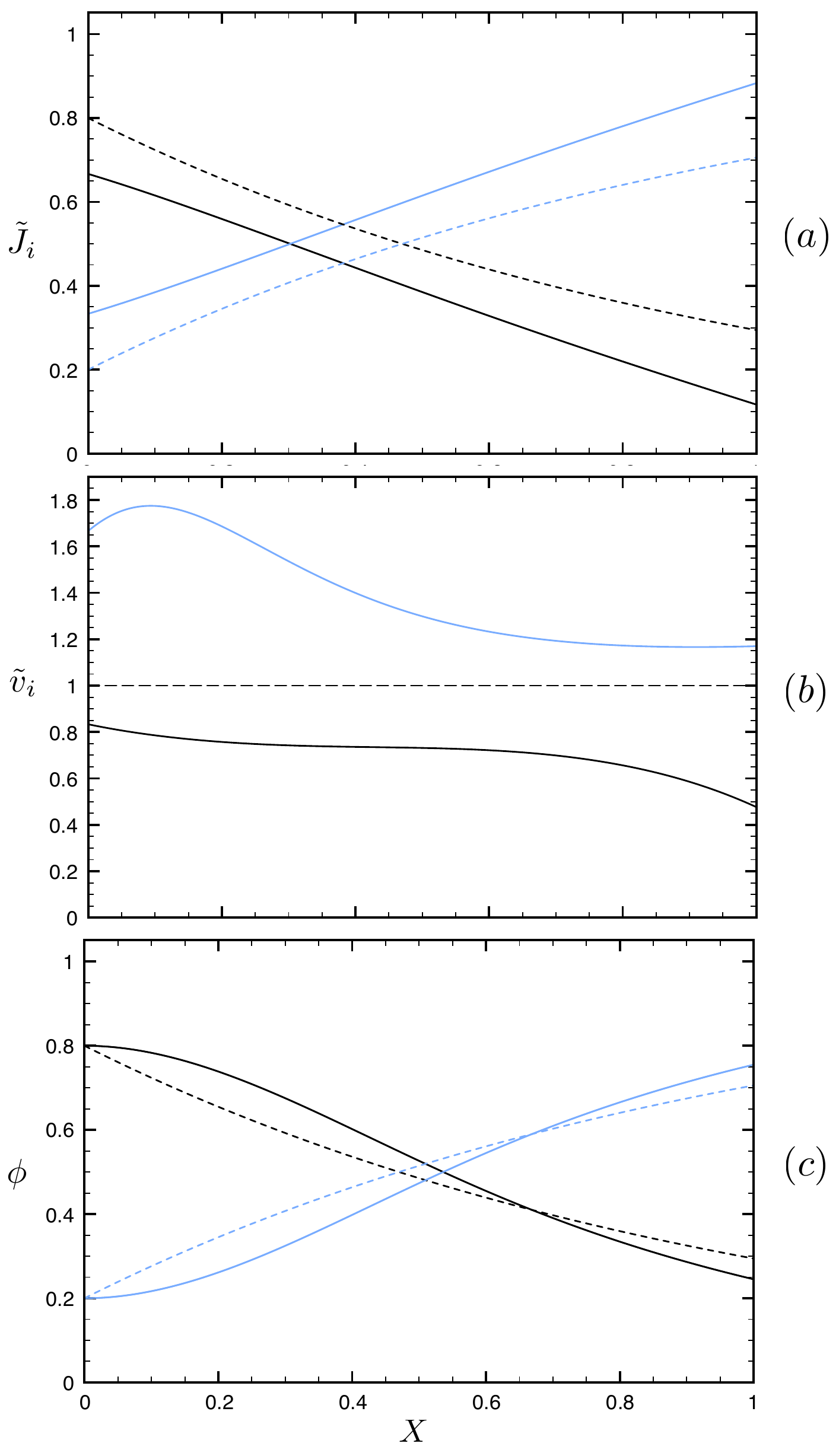}
\caption{(a) Dimensionless currents $\tilde{J_i}$, (b) dimensionless velocities $\tilde{v}_i$, and (c) volume fraction $\phi$,  as a function of the dimensionless position $X$ in the reactor, obtained from numerically solving Eq.~\eqref{ConsNDpsi}, with $\xi_1=\xi_2=1$. The parameters $\tilde{\Lambda}_0$, $\alpha_\text A$, $\alpha_\text B $ and $r$ are set to 1. Black corresponds to species A and blue to species B. Dashed lines signifies the monophasic case with chemistry.}
\label{phiJV}
\end{center}
\end{figure}

A better understanding of the profile shapes can be gained from the examination of the spatial  evolution of the normalized currents $\tilde{J_i} = J_i / J$, and normalized speed defined as $\tilde{v}_i = \tilde{J_i}/\phi_i$. Their evolution with the position is shown in Fig.~\ref{phiJV} where both $\xi_1$ and $\xi_2$ are set to 1. For comparison, each graph also displays the profiles in the monophasic case. One observes on Fig.~\ref{phiJV}(a) that the currents in the two-phase case are not significantly modified when comparing to the monophasic situation, apart from a slowing down, respectively an increase, of the current for the wetting reactant phase, respectively the non-wetting product phase. 
By contrast, Fig.~\ref{phiJV}(b) reveals that the true difference lies in the relative speed of each species. While in the monophasic case the velocity is constant, the velocities of the two phases,  in the biphasic situation, exhibit very different and non-trivial behavior. The most remarkable one is that of the product B phase, which shows a well marked maximum near the reactor inlet. 
The intuition for such a behavior is the following. Close to the entrance of the reactor, the fractions of the two species are very different and cannot vary rapidly because of the boundary conditions at $X=0$, which not only fix their values but also impose a slow spatial variation through $\partial_X \psi = 0$. The current variations, which are identical in absolute values, thus principally translate into velocity variations, the magnitudes of which are pondered by the volume fractions of the phases. Namely, the spatial  increase rate of the velocity of phase B, must be $\phi_\text A/\phi_\text B$ times larger than the spatial decrease rate of the velocity of phase A. This regime is maintained until the variations of the volume  fractions start 
increasing in magnitude, which takes place on the scale \textcolor{black}{$\sqrt{A(\psi) \xi_1 \xi_2}$, given by} the amplitude of the second order derivative term of  Eq.~\eqref{ConsNDpsi}. Further downstream, the variations of the phase fractions become significant. Since the velocity variation of phase A remains small, one obtains that the sign of the velocity variation of phase B is given by the sign of $\tilde{v}_A - \tilde{v}_B < 0$.  The velocity of phase B decreases, which explains the non-monotonic behavior reported above. 
Note that the above mechanism is not based on which species is wetting or non-wetting, which \textcolor{black}{has} also been confirmed numerically (not shown here).

These non-monotonous behaviors of both the velocity of phase B and the spatial transformation rate, that is the gradient of volume fraction along the reactor, are non-trivial, and yield qualitatively significant effects.  In particular, the position of the inflection point depends on $\xi_1$ and $\xi_2$, opening the way towards optimization strategies. Before providing an example of such a strategy, let us first better characterize the respective roles of $\xi_1$ and $\xi_2$. Increasing both $\xi_1$ and $\xi_2$ lead to an overall spatial slowing down of the chemical transformation, although in very different ways. Tuning $\xi_1$ has a strong impact on the shape of the profile, while $\xi_2$ essentially rescales, albeit not strictly speaking, the length on which the chemical transformation occurs.
\begin{figure}[t!]
\begin{center}
\includegraphics[,scale=0.45]{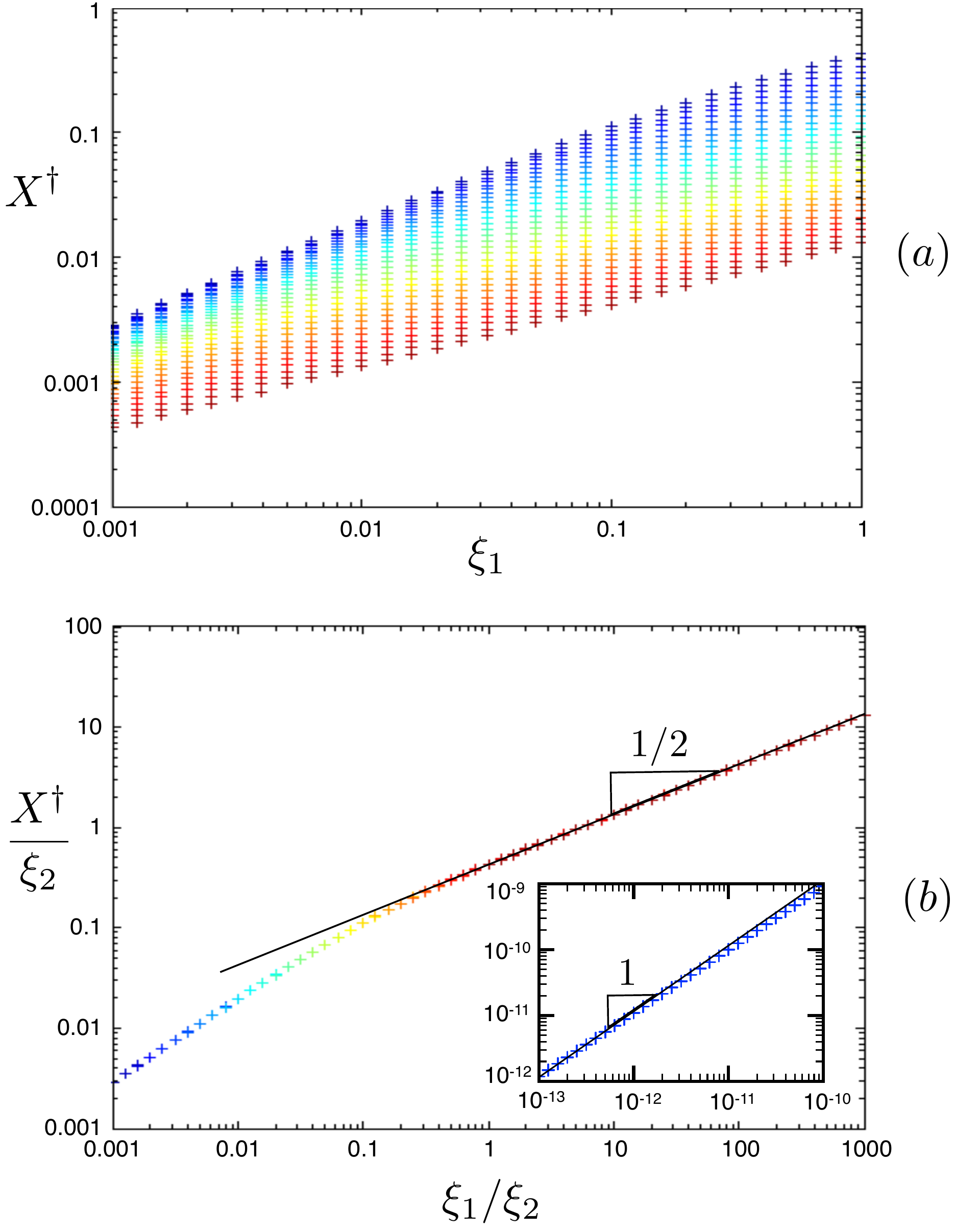}
\caption{(a) Position of the inflection point $X^{\dagger}$ \textcolor{black}{of the curve $\psi(X)$ (see Fig.~\ref{solutions1})} along the reactor as a function of the dimensionless parameter $\xi_1$ for a set of \textcolor{black}{31} logarithmically-spaced values of $\xi_2\in [0.001,1]$ (from red to blue). (b) Rescaled version of the same data set when plotting $X^{\dagger}/\xi_2$ as a function of the ratio $\xi_1/\xi_2$ (with the same color code). 
The parameters $\tilde{\Lambda}_0$, $\alpha_\text A$, $\alpha_\text B $ and $r$ are set to 1. \textcolor{black}{The inset corresponds to very small values of $\xi_1/\xi_2$.}}
\label{scaling}
\end{center}
\end{figure}
Figure~\ref{scaling}(a) displays the position $X^{\dagger}$ of the inflection point along the reactor, as a function of $\xi_1$, for a set of values of $\xi_2$. As observed on the profiles (see Fig.~\ref{solutions1}), both the increase of $\xi_1$ and $\xi_2$ lead to a downstream displacement of the inflection point. More spectacular is that this set of data can actually be rescaled (see Fig.~\ref{scaling}(b)) according to:
\begin{equation}
\frac{X^{\dagger}}{\xi_2} = h(\xi_1/\xi_2) \, , \label{Xdag} 
\end{equation}
\textcolor{black}{with the following scalings:}
\begin{subequations}
\begin{align}
&h(u)  \sim \textcolor{black}{u}
 \,\,\,\,\,\,\,\,\,\,\,\,\,\,\,\,\,\,\, \text{when} \, u \,\,  \textcolor{black}{\ll 1}
  \ \label{scal1}\\
&h(u)  \sim u^{1/2} \,\,\,\,\,\,\,\,\,\,\,\, \text{when} \, u\gtrsim1.  \label{scal2}
\end{align}
\label{eq:scaling}
\end{subequations}
One distinguishes two regimes governed by the intrinsic dimensionless number, 
$\frac{\xi_2}{\xi_1} = \textrm{Ca} \frac{a^2}{\kappa} \frac{l_\chi}{a}$, already introduced in Eq.~\eqref{beta}, which does not depend on the reactor length. For small $\frac{\xi_2}{\xi_1}$, one has a simple scaling $X^{\dagger}\sim (\xi_1\xi_2)^{1/2}$, which depends solely on the product $\xi_1\xi_2$, as suggested by the balance of the higher order derivative term of Eq.~\eqref{ConsNDpsi} and the chemistry term. On the contrary, when the chemical length scale becomes significant, that is for large $\frac{\xi_2}{\xi_1}$, one obtains \textcolor{black}{$X^{\dagger}\sim \xi_1$. Interestingly, both scalings can be found analytically by renormalizing $X$, respectively by $(\xi_1\xi_2)^{1/2}$ and $\xi_1$, in the governing Eq.~\eqref{ConsNDpsi}. By doing so, we are able, in both asymptotic regimes, to eliminate the dependencies of the governing Eq.~\eqref{ConsNDpsi} on $\xi_1$ and $\xi_2$. This actually means that, in these extreme regimes, not only $X^{\dagger}$ becomes $\xi_1$ and $\xi_2$ independent, but the entire profiles actually rescale when plotting the volume fraction as a function of the relevant variables, namely $X/(\xi_1\xi_2)^{1/2}$ and $X/\xi_1$.} 
 
\subsection*{Optimization}

Finally, we would like to provide a simple example of optimization scenario for the type of chemical porous reactors we have considered. The purpose of this part is of course not to present a realistic optimization process, which, for a real system, would take place in a higher dimensional parameter space. Rather, we would like to illustrate how in this simplest case, the identification of the relevant dimensionless control parameters, together with their dependencies on the physical quantities indeed provide a natural route towards optimization.

\begin{figure}[t!]
\begin{center}
\includegraphics[,scale=0.47]{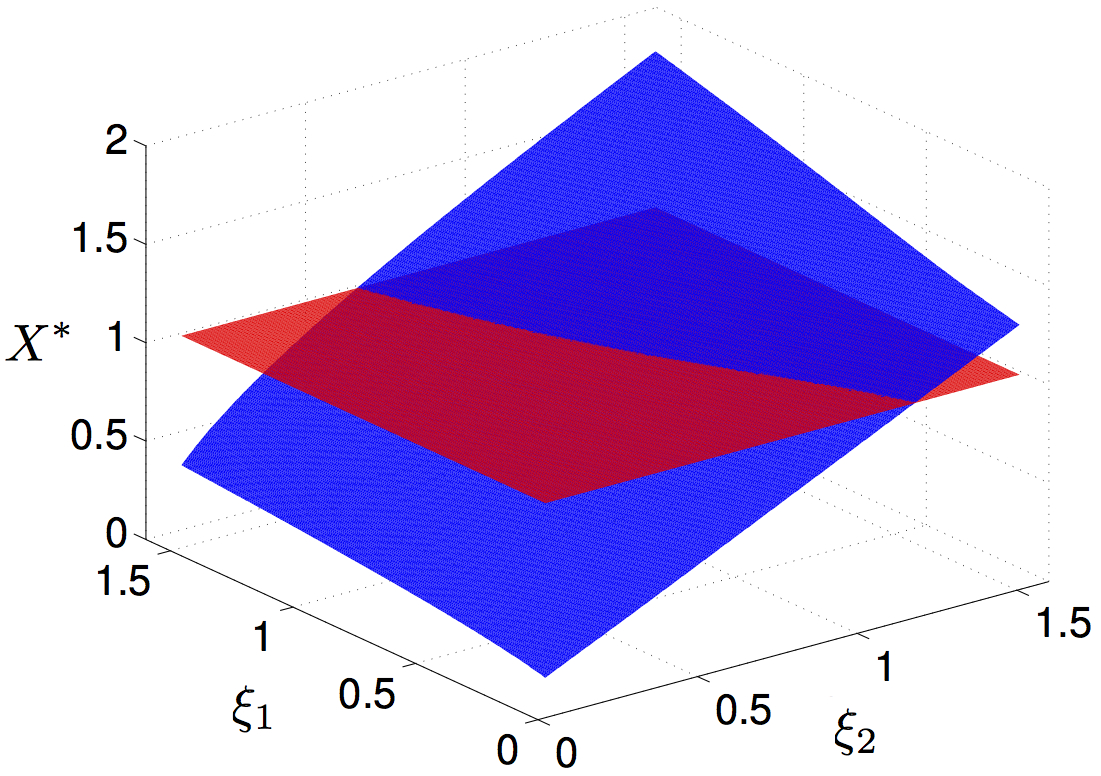}
\caption{Dimensionless position $X^*$ defined through the equation $\psi(X^*)=0$ (blue), and the $X^*=1$ plane (red) against the dimensionless lengths $\xi_1$ and $\xi_2$. The parameters $\tilde{\Lambda}_0$, $\alpha_\text A$ $\alpha_\text B $ and $r$ are set to 1.}
\label{Nappe}
\end{center}
\end{figure}

\begin{figure}[t!]
\begin{center}
\includegraphics[,scale=0.5]{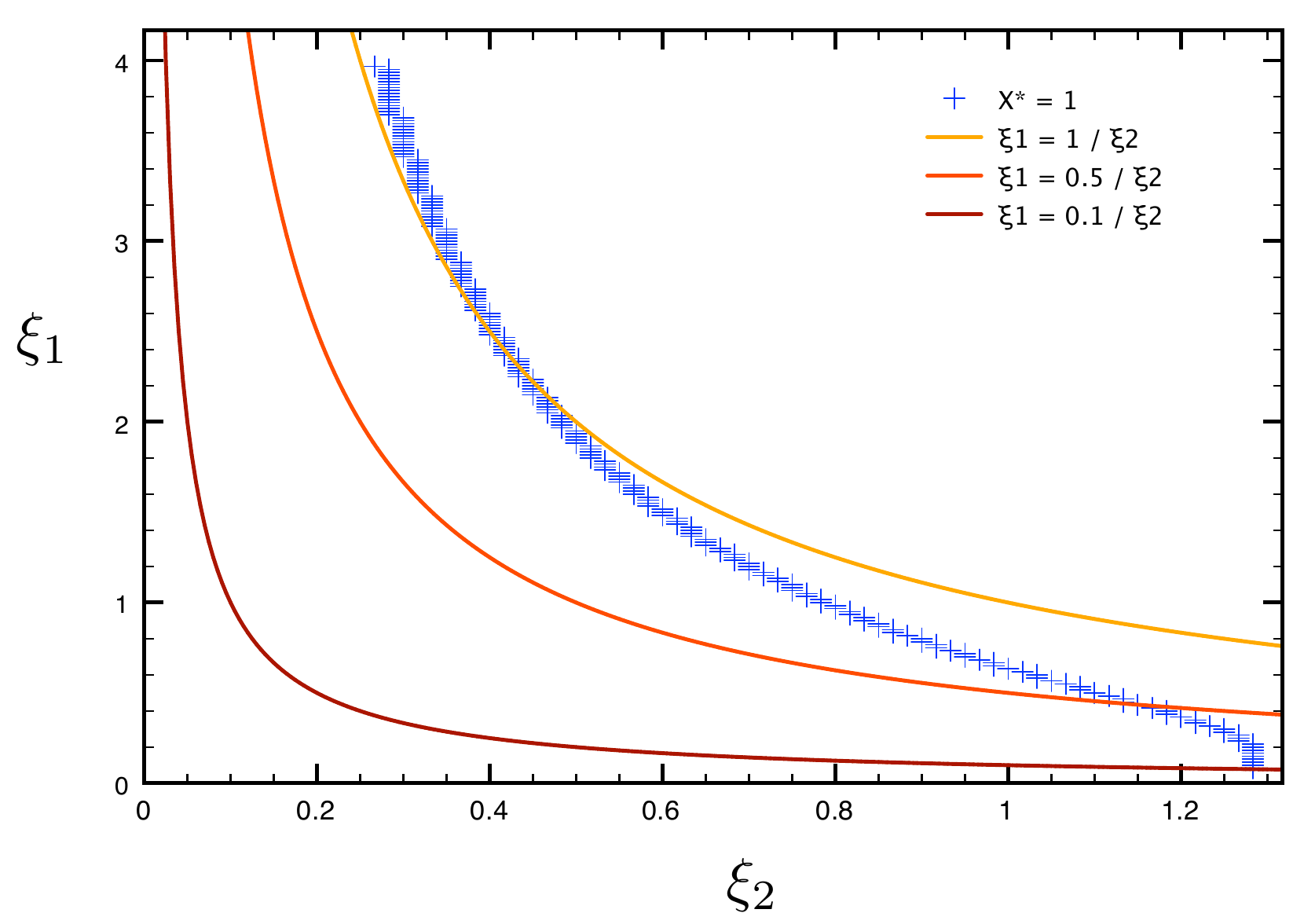}
\caption{Numerical optimal curve $\xi_1=g(\xi_2)$ (blue crosses) and parametric plots (solid lines) of the dimensionless parameter $\xi_1$ against the dimensionless parameter $\xi_2$ as given by Eq.~\eqref{xis}, for different reactor lengths $L$ (the other physical quantities in Eq.~\eqref{xis}  being unchanged). The parameters $\tilde{\Lambda}_0$, $\alpha_\text A$, $\alpha_\text B $ and $r$ are set to 1.}
\label{Hyperboles}
\end{center}
\end{figure}

Let us first recall how $\xi_1$ and $\xi_2$ depend on the physical parameters and rewrite Eqs.~\eqref{xi11} and \eqref{xi22} as:
\begin{subequations}
\begin{align}
\xi_1&= \frac{\gamma \kappa}{\mu a J L}\, \label{XI-1} \\
\xi_2&= \frac{J\tau}{L}\, . \label{XI-2}
\end{align}
\label{params}%
\end{subequations}
From these expressions, it is clear that $\xi_1$ and $\xi_2$ can be tuned independently, playing only with the total flow rate $J$ and the length of the reactor $L$, which does not  require any fine adjustment of the fluid properties and/or the raw permeability of the reactor. To proceed further, some kind of cost function must be defined. Let us choose for instance that the quantity to maximize is the volume fraction of product B at the exit of the reactor.  Namely, if $X^*$ 
denotes the dimensionless position such that $\psi(X^*)=0$, one would like to set ($\xi_1,\xi_2$), such that $X^*=1$. 
The blue surface on Fig.~\ref{Nappe} shows the dependence of $X^*$ on $\xi_1$ and $\xi_2$. 
Intersecting this surface with the plane of equation $X^*=1$ leads to the numerical optimal curve $\xi_1=g(\xi_2)$, displayed with blue crosses on Fig.~\ref{Hyperboles}.
 
As a concrete example, let us now consider the situation where the length of the reactor $L$ is fixed, and see how to identify the best flow rate $J$. Eliminating the flow rate $J$ in Eqs.~\eqref{params},  $\xi_1$ can be expressed as a function of $\xi_2$ as:
\begin{eqnarray}
\xi_1&=&\frac{\gamma \kappa \tau}{a \mu L^2} \frac{1}{\xi_2} \, . \label{xis}
\end{eqnarray}

\noindent Figure~\ref{Hyperboles} displays the evolution of $\xi_1$ as a function of $\xi_2$, for different reactor lengths $L$ (the other physical quantities in Eq.~\eqref{xis} being unchanged). According to Eq.~\eqref{params}, varying the current now corresponds to following the curve corresponding to the desired fixed length of the reactor. The intersection with the numerical optimal curve $\xi_1=g(\xi_2)$ provides the optimal values $\xi_1^{\text{opt}}$ and $\xi_2^{\text{opt}}$ from which the optimal flow rate is obtained:
\begin{eqnarray}
J_{\text{opt}}&=&\frac{\xi_2^{\text{opt}} L}{\tau} \, . \label{Jopt}
\end{eqnarray}
Using similar arguments, one easily finds the corresponding expression for the length $L$ if, instead of considering a fixed reactor length, one had fixed the current and optimized the reactor length. \textcolor{black}{As mentioned previously, the optimization can be approached from a different perspective, and one could aim at maximizing the current $J_\text B$ rather than maximizing its volume fraction. Same kind of reasoning could thus be used and would yield different sets of optimized parameters.}

\begin{figure}[t!]
\begin{center}
\includegraphics[,scale=0.46]{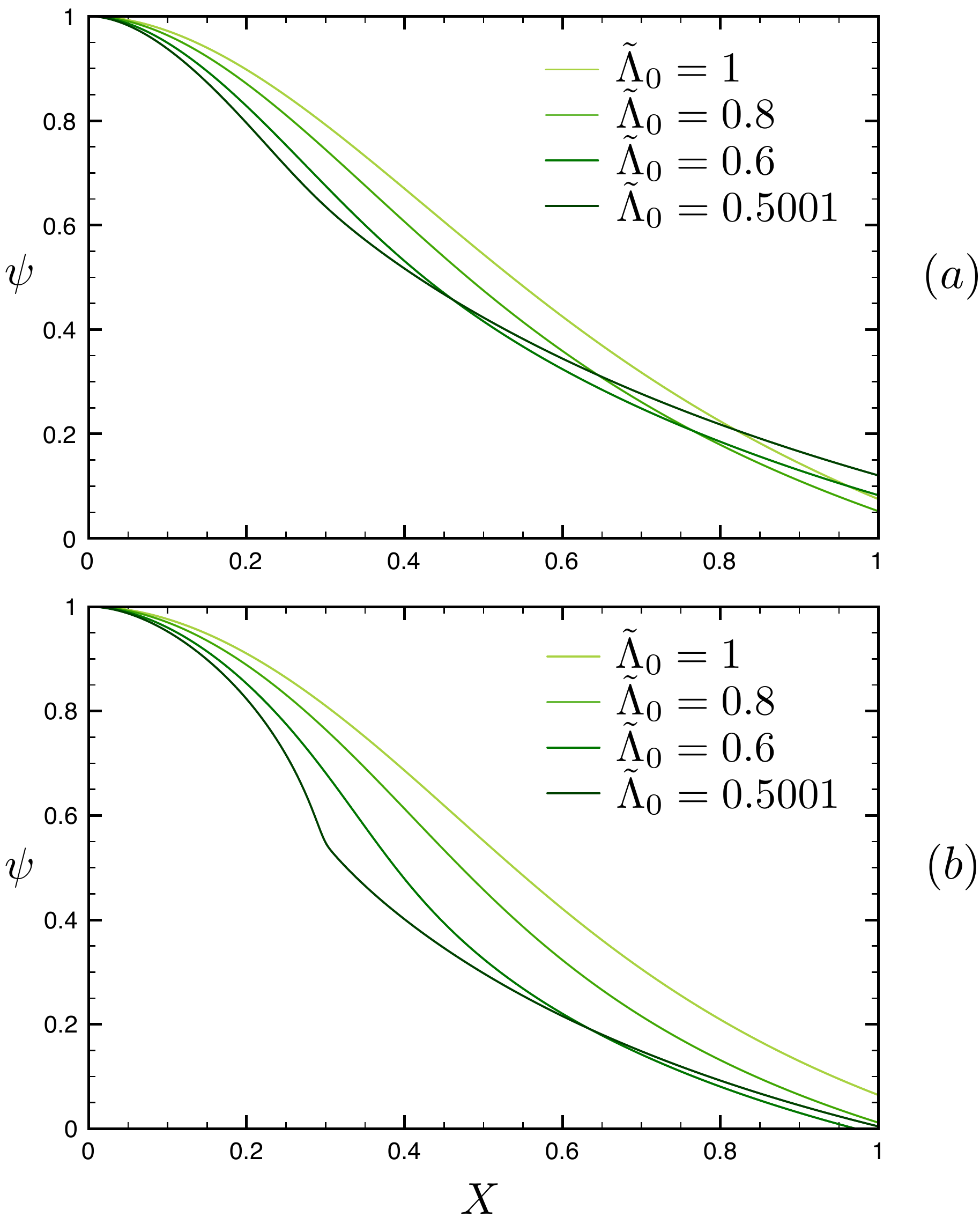}
\caption{Renormalized volume fraction $\psi$ against the dimensionless position X in the reactor for different values of the parameter $\tilde\Lambda_0$, which controls the approach to the solvability condition criterion given by Eq.~\eqref{antiONS2}. The parameters $\xi_1$, $\xi_2$, $\alpha_\text A$, $\alpha_\text B $ are set to 1. (a) r=1 and (b) r=0.25.
} \label{critical}
\end{center}
\end{figure}

Lastly, let us emphasize that all the phenomenology that we have described, in particular the non-monotonous behavior of the spatial transformation rate, sharpens when the system approaches criticality in the sense of satisfying the solvability condition~\eqref{antiONS} through its limiting inequality. This is illustrated on Fig.~(\ref{critical}), where a set of volume fraction profiles obtained with $\xi_1 = \xi_2 = 1$ are displayed for different values of $\tilde\Lambda_0$, which set the distance to criticality according to Eq.~\eqref{antiONS2}. For the present set of parameters, the critical value of $\tilde\Lambda_0$ is $0.5$ (see Fig.~\ref{critical}(a))
. The closer to the critical situation, the more the inflection point moves upward in the reactor. Note that this effect can be all the more dramatic when changing the set of parameters (see Fig.~\ref{critical}(b)).

The above observation has two implications. Firstly, it means that the spatial extent of the region where most of the chemical transformation occurs reduces when approaching criticality. This may have several consequences of practical concern, through for instance a strong localization of the pollution of the catalyzer. Secondly, it opens the way towards more effective optimization strategies.  It would thus be of interest to control the distance to criticality. However, in contrast with the dimensionless lengths $\xi_1$ and $\xi_2$, which can be tuned in independent ways rather easily, the relative permeabilities and the viscous couplings strongly depend on the inner details of the reactor. Optimizing the criticality would thus require a much better knowledge of the functional dependencies of these quantities with the volume fraction $\phi$. Up to now, these functions are essentially prescribed phenomenologically. The present results suggest that there is room for a significant improve of chemical reactors efficiency, and therefore call for more studies of the mechanisms at play at the pore scale.

\section*{Conclusion}

We presented a seminal theoretical approach to a problem in which two-phase flows in porous media are combined with chemistry. Considering a spontaneous first order chemical reaction A $\rightarrow$ B, where A and B are two immiscible species that flow into a porous reactor, we derived the evolution equation for the volume fraction. We also presented a simple model for the average capillary pressure and discussed the importance taken by the viscous coupling terms in the extended Darcy's law. We notably showed that they must satisfy a certain inequality for the problem to be solvable. After restricting our description to a range of volume fractions, where the system can be well modeled, we solved numerically the governing equation in the stationary regime and discussed the behavior of the volume fraction as a function of distance along the reactor when varying $\xi_1$ and $\xi_2$, the relevant dimensionless parameters in this problem. We highlighted the fact that the study of two-phase flows in porous media with chemistry cannot be seen as a simple interpolation of two limit cases, namely monophasic with chemistry and biphasic without chemistry. In particular, we observed that the spatial transformation rate along the reactor is not monotonous. There is a distinct region of the reactor where the chemical transformation concentrates. We also obtained the scalings for the location of this region with $\xi_1$ and $\xi_2$. We finally illustrated how the present kind of analysis can be used for a basic but efficient first level optimization of a given reactor, which does not require fine tuning of its inner characteristics. Of course, the present study calls for experimental validation. We propose precise scaling relations which could and should be tested. Detailed experimental studies are also necessary to constrain the modelization of the Darcy coefficients. Our best wish is that the present work contributes to \textcolor{black}{motivating} 
such studies. 

\section*{Acknowledgements}
We thank J.~P. Dath from Total Petrochemicals Research Feluy for having attracted our attention on this very interesting problem. We acknowledge S. Datta, D. Lacoste, \textcolor{black}{Y. Meheust}, E. Rapha{\"e}l, B. Semin and C. Souli{\'e} for fruitful discussions. 

\bibliographystyle{h-physrev}
\bibliography{biblio}

\end{document}